# Categorical Judgment with a Variable Decision Rule

Burton S. Rosner and Greg Kochanski

*University of Oxford*

# Abstract

A new Thurstonian rating scale model uses a variable decision rule (VDR) that incorporates three previously formulated, distinct decision rules.  The model includes probabilities for choosing each rule, along with Gaussian representation and criterion densities. Numerical optimisation techniques were validated through demonstrating that the model fits simulated data tightly. For simulations with 400 trials per stimulus (*tps*), useful information emerged about the generating parameters.  However, larger experiments (e.g. 4000 *tps*) proved desirable for better recovery of generating parameters and to support trustworthy choices between competing models by the Akaike Information Criterion. In reanalyses of experiments by others, the VDR model explained most of the data better than did classical signal detection theory models.

Imagine conducting a rating scale experiment with $N$ stimuli that vary on some property. You present the subject with a single randomly selected stimulus $S_h$ on each trial. The subject rates it on a scale isomorphic with the integers 1, …, $M$. Typically, each $S_h$ occurs on multiple trials. The resulting data form an $N$ by $M$ matrix **N** where cell $(h,i)$ gives the conditional number of trials $n(\mathrm{R}=i \,|\, S_h)$ on which stimulus $S_h$ evoked response $\mathrm{R}_i$.

The standard Thurstonian model for a rating experiment asserts that the momentary representations $s_{ht}$ of each stimulus $S_h$ across trials $t$ form a Gaussian density on a unidimensional psychological continuum $v$. In the model, the subject also has $M$-1 criteria $C_i$ that divide $v$ into $M$ regions, and the rating given to a $s_{ht}$ depends on which region of the continuum it occupies. Criterion positions vary randomly across trials, independently of the representations and of each other. Each $C_i$ generates a Gaussian density of positions on $v$.

As a result of these assumptions, on any trial $t$ the observer has a representation $s_t$ and $M$-1 ordered criterion samples $c_{it}$, $i = 1, …, M$-1 that partition $v$. The representation and criterion samples are the only basis of the observer's decision; notably, the observer does not know which $S_h$ gave rise to $s_t$. To arrive at a response, the observer must apply some decision rule to these momentary subjective events. The rule depends on the relationship between $s_t$ and the $c_{it}$.

The long accepted mathematical form for this model, the Law of Categorical Judgment (Torgerson, 1958; Wickelgren, 1968; McNicol, 1972) proved to be wrong. Rosner and Kochanski (2009) demonstrated that the equation could produce negative conditional response probabilities $p(\mathrm{R}=i \,|\, S_h)$ when criteria had nonzero variances and varied independently. Those authors presented a new equation, the Law of Categorical Judgment (Corrected), that mended the fault. Reversing the original order of their first two integrations gives this equivalent and computationally more tractable form:

$$p(\text{R} = i \,|\, \text{S}_h) = \int_{-\infty}^{\infty} \phi(s_h; \mu_{S_h}, \sigma_{S_h}) \int_{s_h}^{\infty} \phi(c_i; \mu_{C_i}, \sigma_{C_i}) \prod_{j \neq i}^{M-1} [1 - \int_{s_h}^{c_j} \phi(c_j; \mu_{C_j}, \sigma_{C_j}) \, dc_j] \, dc_i ds_h, \quad (1)$$

where $\phi$ is a Gaussian density.

Equation 1 rests on a particular decision rule, designated Rule 1. Under it, the observer computes the $M$-1 momentary differences ($c_{kt} - s_t$) between the representation $s_t$ and each criterion sample $c_{kt}$. If ($c_{kt} - s_t$) is the smallest positive difference, the observer makes response $K$; if all differences are negative, the default response is $M$. (L.L. Thurstone offered this rule in his lectures on psychophysics in 1947, which one of us attended.)

Klauer and Kellen (2010) then pointed out that Rule 1 is not the only possibility. Moreover, they showed that different decision rules can have observably different consequences.  For instance, they demonstrated that Rule 1 biases responses towards high values. To do this, they considered a rating experiment where the parameters are symmetric: representation and criterion means are evenly spaced; the middle criterion mean coincides with the middle representation mean; and all variances are equal. Under these symmetric assumptions, one might expect cells ($i$+$k$, $j$+$l$) and ($N$-$k$, $M$-$l$) in the data matrix **N** to contain equal entries. Klauer and Kellen established that Rule 1 instead produces asymmetric results even though the underlying parameters are symmetric.

This observation spurred Klauer and Kellen (2010) to offer a second decision rule that generates the opposite asymmetry.  Under their Rule 2, the observer computes the $M$-1 momentary differences ($s_t - c_{it}$). If ($s_t - c_{itk}$) is the smallest positive difference, the observer makes response ($K$+1). If all differences are negative, the default response is 1. Rule 2 biases responses towards low values. Its equation is

$$p(\text{R} = i+1 \,|\, \text{S}_h) = \int_{-\infty}^{\infty} \phi(s_h; \mu_{S_h}, \sigma_{S_h}) \int_{-\infty}^{s_h} \phi(c_i; \mu_{C_i}, \sigma_{C_i}) \prod_{j \neq i}^{M-1} [1 - \int_{c_j}^{s_h} \phi(c_j; \mu_{C_j}, \sigma_{C_j}) \, dc_j] \, dc_i ds_h, \quad (2)$$

Klauer and Kellen (2010) then proposed another rule, Rule 3, based on the absolute differences $|s_t - c_{it}|$. This rule yields symmetric results when the underlying parameters are symmetric. The equation for Rule 3 has three parts:

a) $p(\text{R}=1|\text{S}_i) = \int_{-\infty}^{\infty} \phi(s_h;\mu_{S_h},\sigma_{S_h}) \int_{s_h}^{\infty} \phi(c_1;\mu_{C_1},\sigma_{C_1}) \prod_{j\neq 1}^{M} [1 - \int_{2s_h-c_1}^{c_1} \phi(c_j;\mu_{C_j},\sigma_{C_j})\, dc_j]\, dc_1 ds_h,$

b) $\quad p(\text{R}=i|\text{S}_h) = \int_{-\infty}^{\infty} \phi(s_h;\mu_{S_h},\sigma_{S_h}) \int_{s_h}^{\infty} \phi(c_i;\mu_{C_i},\sigma_{C_i}) \prod_{j\neq i}^{M} [1 - \int_{2s_h-c_i}^{c_i} \phi(c_j;\mu_{C_j},\sigma_{C_j})\, dc_j]\, dc_i ds_h$

$\qquad + \int_{-\infty}^{\infty} \phi(s_h;\mu_{S_h},\sigma_{S_h}) \int_{-\infty}^{s_h} \phi(c_{i-1};\mu_{C_{i-1}},\sigma_{C_{i-1}}) \prod_{k\neq i-1}^{M} [1 - \int_{c_{i-1}}^{2s_h-c_{j-1}} \phi(c_j;\mu_{C_j},\sigma_{C_j})\, dc_j]\, dc_{i-1} ds_h,$

$\qquad i = 2,...,M-1,$ and

c) $\quad p(R=M\,|\,S_i) = 1 - \sum_{j=1}^{M-1} p(R=j\,|\,S_i)$ \hfill (3).

(The authors gave a computationally more complicated version of part c.)

The heart of Klauer and Kellen's (2010) analysis is that opposite asymmetries occur under Rules 1 and 2 while no asymmetries arise under Rule 3. The authors urged experimenters using the rating scale method to "check their major conclusions for robustness under different choices of decision rule".

Cabrera, Liu, and Dosher (2015) used the three decision rules separately in modelling a multipass category rating (MCR) detection experiment. They generated pseudo-data with their model, using one choice of rule. Then they fitted the model using that rule or a different one. The best fits occurred when the generating and fitted rules coincided. Cabrera et al. also found that their model using Rule 1 gave the best fit to MCR visual detection data that they collected.

Klauer and Kellen (2010) pointed out, however, that observers could adopt probabilistic mixtures of decision rules when criteria can change ordinal positions. Therefore, an infinite number of decision procedures are available. But how then can experimenters check their conclusions for robustness, given this universe of possible decision procedures? Carbrera et al. (2015) were aware of this question and promised to deal with it in the future.

We now provide a straightforward, systematic answer to the question.  Equations 1, 2, and 3 must be fitted to data by numerical optimisation. Since Klauer and Kellen's (2010) multiple decision procedures are probabilistic mixtures of rules 1, 2, and 3, we simply include those probabilities as mew parameters in our numerical optimisation.

The observer chooses probabilistically between Rules 1, 2, and 3 on each trial, and the probabilities of selecting the rules are parameters of our new model. They are optimised numerically, along with the parameters of the Gaussian representation and criterion densities. This variable decision rule (VDR) model naturally handles data generated by Rules 1, 2, or 3 individually or in any probabilistic combination. We believe this approach covers all reasonable decision procedures built directly on top of independent normal criterion densities.

After firstly specifying the VDR model, we secondly report tests on pseudo-data generated by a variety of models with different decision conditions. The tests examine the full VDR model, restricted versions of it, and special cases with zero criterion variance or zero signal variance. The Akaike information criterion (corrected) (cf. Burnham & Anderson, 2004) is used to select the best model where required. Thirdly and finally, we apply our VDR model to a pair of published experiments.

**Variable decision rule (VDR) model**

The VDR rating scale model assumes the usual independent Gaussian densities for the $N$ stimulus representations and the $M$-1 criteria. Each density has its own mean and standard deviation. These parameters are subject to certain conditions. The representation means must always be in ascending order, as must the criterion means.

We now introduce three new parameters, $p_{R1}$, $p_{R2}$, and $p_{R3}$. These $p_{Rk}$ are the probabilities that across trials the observer chooses Rules 1, 2, or 3, respectively. They are optimised along with the representation and criterion density parameters. Like any other parameter in the VDR model, each $p_{Rk}$ can be unconstrained or constrained, subject however to two conditions. First, each $p_{Rk}$ must lie in the closed interval [0,1]. Second, the three probabilities must always add to unity. Therefore, at most two probabilities can be independently adjusted. Computationally, a matrix $\mathbf{P}_k$ of conditional probabilities $p(R = i|S_h)$ is obtained for each rule, given a set of signal and criterion density parameters. Each $\mathbf{P}_k$ is multiplied by rule probability $p_{Rk}$. Then a final conditional probability matrix $\mathbf{P}$ results from $\mathbf{P} = \sum_k p_k * \mathbf{P}_k$ .

The *general* VDR model uses a probabilistic mixture of all three decision rules, represented as a set $DR = \{1, 2, 3\}$ where $p_{Rk} > 0$ for $k \in DR$.  There are six restricted classes of VDR models. In three dual-rule models, one $p_{Rk}$ is zero so that Rule $k$ is not used. The set $DR$ contains only two members, e.g., $DR = \{1, 3\}$ where Rule 2 never applies. Finally, in each of the three single-rule models, one $p_{Rk}$ is unity and the other two are zero. The rule set becomes $DR = \{k\}$. This gives a total of seven classes of VDR models.

A data matrix has $N(M\text{-}1)$ degrees of freedom. The VDR model requires estimating $U = [2N + 2(M\text{-}1) + k]$ parameters, where $0 \leq k \leq 2$ is the number of free rule probabilities. To be able to solve the equations, we must have more equations (i.e. elements of the data matrix) than model parameters. Thus, a necessary condition for fitting a VDR model is $N(M\text{-}1) > U$. To solve these equations, one's experiment needs at least N$\geq$3 and M$\geq$4, and it must not be too close to (M, N) = (0, 0).

In the VDR model, we make no assumptions about representation and criterion variances. Each variance can have a different value. Accordingly, we do not pay particular attention to constrained forms of the model where representation or the criterion variances are nonzero but are forced to be equal.

Given a set of parameters for the representation and criterion densities, suppose we set $p_{R1} = p_{R2} = 0.5$ and compute $\mathbf{P}_{(1,2)}$. The distribution of nonzero entries along each row of $\mathbf{P}_{(1,2)}$ should be symmetric, allowing for end-effects. Then $\mathbf{P}_{(1,2)}$ might be virtually indistinguishable from the $\mathbf{P}_{(3)}$ computed from the same set of parameters but with Rule 3 alone ($p_{R3}$=1.0). Corresponding entries in cell ($i$, $j$) in the two matrices might always be close. This might make Rule 3 practically redundant, even though Equation 3 clearly cannot be derived as one-half the sum of Equations 1 and 2.

To test this argument, we drew up 10 sets of parameters for five representation densities and nine criterion densities. We combined each set of parameters $l$ with ($p_{R1} = p_{R2} = 0.5$) or ($p_{R3}$=1.0) and computed 10 pairs of 5×10 matrices $\mathbf{P}_{((1,2),l)}$ and  $\mathbf{P}_{((3),l)}$.  Within each pair, we examined corresponding row densities of nonzero conditional probabilities.

The corresponding row densities were roughly similar but displayed some differences. In about one-fifth of the pairs, the maxima occurred at different places. In another fifth, the maxima coincided but differed by .04 or more, flattening one member of the pair. We conclude that the decision procedure ($p_{R1} = p_{R2} = 0.5$) does not computationally duplicate the procedure ($p_{R3}$=1.0). Therefore, we retained Rule 3 in our model. Nevertheless, the small size of the differences will make it hard for the model to distinguish ($p_{R1} = p_{R2} = 0.5$) from Rule 3 alone, as against, say, distinguishing Rule 1 from Rule 2.

*Special cases.* The signal detection theory (SDT) rating scale model has no criterion variance whatsoever (cf McNicol, 1972). Under both the SDT equal-variance (SDT-EV) and the SDT unequal-variance (SDT-UV) forms of the model, the product terms in Equations 1, 2, and 3 are always unity. Both forms are special cases of the VDR model where each of the three decision rules gives the same result. Only a single decision rule is needed; we compute with Rule 1 and use standard routines for obtaining Gaussian tail probabilities.

Another special case of the VDR model has zero representation variance. Each representation density is a Dirac pulse. This is the Complementary Signal Detection Theory (CSDT) model (Rosner & Kochanski. 2010). It includes the three rule probabilities as parameters. The outer integral in Equations 1 through 3 simply drops away; otherwise, the computations are the same as for the VDR model. The CSDT model can adopt one-rule, two-rule, or three-rule forms. Finally, parallel to SDT, it comes in two forms: equal criterion variances (CSDT-EV) and unequal criterion variances (CSDT-UV). Both forms are investigated here.

### Procedures

We next turn to systematic tests of the general VDR model on simulated rating data. Initially, 10 matrices of pseudo-data were generated under each of the seven classes of VDR models. Let $DR_G$ be the set of generating decision rules and $DR_F$ the set of fitted rules. We then fitted the general VDR model ($DR_F=\{1,2,3\}$) and a model of the same class ($DR_F=DR_G$) to each VDR-generated matrix. We examined both the reproducibility and the goodness of each fit. The goodness of fit showed to what extent the recovered parameters matched those underlying the pseudo-data. Then we identified the better of the two fits to each pseudo-data matrix.

*Simulations*. Each simulation operated on a trial-to-trial basis under a given model. Input to the simulation program[1] includes: the number of stimuli ($N$); the number of responses ($M$); the trials per stimulus ($tps$); the desired number ($nsim$) of pseudo-data matrices; the selected decision rules; and an initial set of generating parameters. The latter comprise means and standard deviations for $N$ Gaussian representation densities and $M$-1 Gaussian criterion densities along with three generating probabilities $p_{RkG}$ ($k=1,2,3$) for the decision rules.

On each trial $t$, a value $s_h$ is randomly selected from generating representation density $h$.  A value $c_j$ is randomly chosen from each generating criterion density. If $p_{1G}$ is nonzero, Rule 1 is applied to the differences between $s_h$ and the $c_j$ values, to yield a response $i$. The appropriate cell is incremented in the frequency matrix $\mathbf{N}_{1G}$ for Rule 1. This routine continues until each stimulus has appeared on $tps$ trials. If $p_{R1G}$ is zero, no trials are simulated nor is $\mathbf{N}_{1G}$ computed. The same routine is executed for Rule 2 and then for Rule

3, producing frequency matrices $\mathbf{N}_{2G}$ and $\mathbf{N}_{3G}$, respectively. Finally, the result of

$\sum_k p_{kG} * \mathbf{N}_{kG}$ is rounded cell-wise to yield a generated pseudo-data matrix $\mathbf{N_G}$.

If *nsim*>1, the program probabilistically produces a new set of generating parameter values from the last set. It sorts the new representation and criterion means into ascending order and makes all variances positive. The new generating rule probabilities $p_{1G}$, $p_{2G}$, and $p_{3G}$ are made positive and normalised to a unit sum. Then the program produces a new pseudo-data matrix. This procedure repeats until *nsim* simulations have been executed.

*Optimisation*. For each pseudo-data matrix, we used Bootstrap Markov chain Monte Carlo (BMCMC) (Kochanski and Rosner, n.d.) to find the best fitting values for the Gaussian density parameters and for the rule probabilities $p_{kF}$. The procedure maximizes the posterior log likelihood $\log(L)_F$ of the fit. The optimised values of the rule probabilities $p_{RkF}$ reveal the most plausible decision process underlying the given pseudo-data matrix.

The BMCMC process is iterative, and each optimisation step yields a set of candidate parameters specifying Gaussian representation and criterion densities and the rule probabilities. The parameters are subjected to the same constraints used in the simulations (see above). Then for each nonzero $p_{kG}$ the optimisation program selects the corresponding Equation 1, 2, or 3 and obtains a probability matrix $\mathbf{P_{R\mathit{k}G}}$ from the candidate parameters. A final probability matrix $\mathbf{P}_G$ results from $\sum_k p_{RkG} * \mathbf{P}_{RkG}$ . The program takes $\mathbf{P}_G$ and computes the posterior log likelihood $\log(L)_F$ that the candidate parameters could have produced the pseudo-data matrix $\mathbf{N}_G$. The simulation program follows the same path to get the prior log likelihood $\log(L)_G$ for each pseudo-data matrix, given the generating parameters. A good fit to a pseudo-data matrix here must yield a $\log(L)_F$ close to the $\log(L)_G$ computed by the simulation program.

The Rule(s) used for a given optimisation form a fitted set $DR_F$. As well as fits with the general VDR model ($DR_F$={1,2,3}), our program[2] also provides fits with just one rule ($DR_F$={$k$}) or a probabilistic combination of any pair of rules ($DR_F$={$k, l$}). It also handles SDT and CSDT models.

*Measures of goodness of fit*. The simulation and the optimisation programs compute five measures of goodness of fit between the predicted conditional probabilities $p_F(R=i\,|\,S_h)$ and the "observed" (generated) conditional proportions $P_G(R=i\,|\,S_h)$. Schunn and Wallach (2005) suggested four of these measures: $r^2$ for the regression of the observed proportions on the predicted probabilities; *rmsd*, the root mean square deviation between the observed proportions and the predicted probabilities; and $b_0$ and $b_1$, the two linear regression coefficients. The 95 per cent confidence limits *CL($b_0$)* and *CL($b_1$)* of the coefficients are calculated from standard formulae (see Schunn and Wallach).

The fifth measure, designated *K-L*, is the Kullback-Leibler coefficient of divergence (Kullback & Leibler, 1951; cf Kullback, 1959). It expresses the relative entropy of two probability densities *P* (data) and *p* (model). Unlike the other measures, *K-L* is sensitive to large relative errors between small probabilities. It is specified here as

$$\sum_h \sum_i P_G(R = i\,|\,S_h)\log_2[P_G(R = i\,|\,S_h) / p_F(R = i\,|\,S_h)] \tag{4}.$$

### Results: VDR-generated pseudo-data

Rating experiments were simulated with *N*=5, *M*=10, and *tps*=400. The value for *tps* comes from rating experiments by Schouten and van Hessen (1998) that exemplify a large but practical study. We experimented with seven groups of VDR models: three groups where one $p_{RkG}$ is nonzero, three more groups where two $p_{RkG}$ are nonzero; and the general VDR model where all three $p_{RkG}$ are nonzero.

Ten sets of generating parameters were produced under each of the seven groups. This gave 70 matrices of pseudo-data. We fitted both the general VDR model ($DR_\mathrm{F}$={1,2,3}) and the generating model ($DR_\mathrm{F} = DR_\mathrm{G}$) to each data matrix. This gave a total of 130 fits (10+2*60). (Note that the general VDR model was also the generating model for 10 matrices.)

*Reproducibility*. This undertaking presents a complicated optimisation problem. We have no proof that our -log($L$) is concave, that is has a single minimum, or that our optimisations have fully converged. The BMCMC algorithm uses techniques analogous to simulated annealing; it may hop from one local minimum to a better one, but it cannot guarantee termination at the global minimum.

Therefore, at best we can only check the reproducibility of any solution for a given pseudo-data matrix by repeating the optimisation on it from different starting points. If the results seem consistent, the optimisation with the highest log($L$) is accepted for further treatment. Accordingly, each optimisation on a pseudo-data matrix was repeated four times. Four sets of initial values were devised for the representation and criterion density parameters. They were used in each set of repeated optimisations. The values of $p_\mathrm{R1F}$, $p_\mathrm{R2F}$, and $p_\mathrm{R3F}$ varied both within and across sets of starting points.

To assess agreement between the optimisations on a pseudo-data matrix, we calculated a per cent measure of inconsistency *%ic* for each set of four repeats:

$$\%ic = 100 * [\max(\log(L)_\mathrm{F} - \min(\log(L)_\mathrm{F}] / [\max(\log(L)_\mathrm{F} + \min(\log(L)_\mathrm{F}] / (-2) \qquad (5).$$

Notice that a low value reflects highly reproducible optimisations and that log($L$) is always negative.

The question now arises whether the fits with the general VDR model are as consistent as the fits of the model that generated the pseudo-data.  We checked this with a Wilcoxon paired samples test on *%ic* values. We computed  *%ic* for the fits of both models to each of the 60 pseudo-data matrices generated through only one or two rules.  The test yielded V=63, P=0.03. The result is that the general VDR fits were marginally less consistent (mean *%ic* = 0.05 vs. 0.04 for the generating model.) If we look at *%ic* for *all* cases where $DR_F = DR_G$, we get a mean *%ic* of 0.05 with a standard deviation of 0.08.

These small values of *%ic* mean that the four solutions for the same pseudo-data matrix had similar likelihoods. This sanctions our taking the best of the four solutions as representative of that set of results and then submitting it to further analysis.

The four solutions in each set always differed slightly from one another. Each usually gave good measures of fit (see below). Therefore, we assume they are close to the global minimum. We can explain these findings as a consequence of our use of Romberg integration (see Press, Teukolsky, Vetterling, & Flannery,  2007) to evaluate Equations 1-3.

Since we have no prior knowledge about the optimal step size for integration, our Romberg integration routine qromo.c starts with a relatively large step size and estimates the resulting integration error.  If the error is too large, qromo.c tries again with a smaller step (1/3 the last size). This step size adjustment repeats until the routine finally gets a small enough error.

This adaptive method works well and guarantees a reasonably small error under almost any conditions.  But it makes discrete changes in step size. This has unexpected consequences because when the step size changes, the computed result of the integral also changes.

Imagine plotting $\log(L)$ as a function of one of the parameters of the VDR model. For example, the first criterion standard deviation could be 0.112, 0.113, 0.114, 0.115, etc, in successive evaluations. Then $\log(L)$ would change in fairly even steps. Now consider one of the integrations inside the product term. The best step size inside that integration may remain at 0.001 over successive criteria. At some point, however, it may switch to 0.003 for the next criterion. The value of that integral, the prediction of the VDR model, and thus log($L$) all suddenly jump. The jump can be upwards or downwards. If log($L$) has been decreasing as the parameter increases, a downward jump has no dramatic effect.  An upward jump, however, creates a spurious local minimum in the log($L$) hypersurface. Furthermore, an upward jump may constitute an unacceptable move for the BMCMC algorithm and require a change of direction in the search.

The VDR model actually requires double integrations.   Consequently, qromo.c is used inside qromo.c itself. The outer qromo.c evaluates the inner one hundreds or thousands of times, and each inner integration separately sets its step size.  We do this double integration for each cell of the data matrix. There are bound to be *many, many* jumps.

Accordingly, the VDR computation is a smooth function as long as integration step size is constant. Jumps occur, however, when one of the step sizes changes. These jumps introduce spurious minima if the smooth trend is downwards and jumps are upwards or vice-versa.  A saw-tooth pattern could result.

The 'temperature' of BMCMC is initially high enough to facilitate escapes from any local minimum. Towards the end of the search, however, as the global minimum draws near, the temperature is low. This makes it harder for BMCMC to escape a local minimum. Thus, a change of integration step size might trap and terminate the search. Optimisations that leave from different starting points are likely to terminate in different, qromo-created local minima near the global minimum. This is why each of our sets of four optimisations yield a cluster of slightly different log($L$) values. Since those values are close, all four optimisations will give about the same (generally acceptable) goodness of fit.

*Goodness of VDR fits*. We first compared log($L$)$_F$ for each best fit to each pseudo-data matrix against the *log(L)*$_G$ computed from the generating parameters.  For this purpose, we computed a per cent difference measure *%$\Delta_{GF}$* for each fit:

$$\%\Delta_{\text{GF}} = 100 * [log(p_{\text{F}}) - log(p_{G})] / [log(p_{\text{F}}) + log(p_{G})] / (-2) \,. \qquad (6)$$

In *all* 130 fits, *%$\Delta_{\text{GF}}$* was positive because $L_{\text{F}}$ exceeded $L_{\text{G}}$.  The fitted parameters always predicted the pseudo-data matrices somewhat better than did the original VDR generating parameters. The mean of *%$\Delta_{\text{GF}}$* was 0.13; the standard deviation was 0.41, dominated by a few large values. (A Wilcoxon test showed no significant difference in *%$\Delta_{\text{GF}}$* between fits with a general VDR model vs. fits with the generating model class; $V = 95$, $p = .78$).

Although initially surprising, a positive *%$\Delta_{\text{GF}}$* is actually the expected outcome for a good fit to a relatively small data set.  Such a set, whether acquired by simulation or by observation, is a snapshot of a statistical fluctuation. Therefore, it represents VDR parameters that are slightly different from the true values used in pseudo-data generation. With only the 400 *tps* used here, the actual pseudo-data frequencies can differ noticeably from their expectation values. As *tps* increases, this discrepancy should decrease, forcing *log(L)$_{\text{F}}$* towards *log(L)$_{\text{G}}$* and reducing  *%$\Delta_{\text{GF}}$*.

The general VDR model had been fitted to 60 pseudo-data matrices generated by dual-rule models and to 60 generated by single-rule VDR models. Of these 120 fits, 118 gave an $r^2$ of at least .95. So did all 10 fits of the general VDR model to psedo-data generated by that model itself. The fits with $r^2 \geq .95$ were accompanied by  *rmsd*$\leq$0.1 and *K-L*$\leq$.05. The confidence intervals for the regression coefficients $b_0$ and $b_1$ always contained 0 and 1, respectively.   All these tight fits demonstrate the power of the BMCMC procedure. They also are to be expected, since we were fitting the generating model and an extended version of it to the pseudo-data. The two remaining cases were exceptional only because $r^2$ was .94. They were the general VDR fit and the fit with $DR_{\text{F}}= DR_{\text{G}}=\{1,3\}$ to the same pseudo-data matrix. That matrix had more small nonzero entries than did the other nine

matrices generated under $DR_G$={1,3}.)

     *Identification of $DR_G$*. Consider the pseudo-data matrices generated under a single-

rule model  ($DR_G$ = {1}, {2}, or {3}). On 20 of the 30 trials, fitting with the general VDR

model ($DR_F$={1,2,3}) correctly identified the largest $p_{kG}$.  This result has a chance probability of $2*10^{-6}$. For the matrices generated under a two-rule model ($DR_G$ = {1, 2}, {1,

3}, and {2, 3}, fitting with the general VDR model correctly identified which $p_{kG}$ was smallest (nearest to zero) in 18 of the 30 attempts. This result is significant at *p*=.0024.

These two sets of findings strongly indicate that fitting with the general VDR model

provides information about the $DR_G$ used in generating the pseudo-data. As the results on

*%Δ*$_{GF}$ imply, of course, the fitted rule probabilities for the general VDR model  do not exactly reproduce their generating counterparts.

     *Model comparisons*. Given a pseudo-data matrix generated by a single-rule or dual-

rule model, that model presumably should fit the pseudo-data better than the general VDR

model. Sixty such matrices had had been fitted with the general VDR model  and with the

VDR model where $DR_F$=$DR_G$.  To compare different models fitted to the same data, our optimisation program computes the Akaike Information Criterion (corrected) (*AICc*).

     In the 60 comparisons, the *AICc* showed that in only 37 instances did the generating

model fit the pseudo-data better than the general VDR model. In a strong minority of 23

cases, the general model proved superior. There were no obvious differences between the results for single- and dual-rule generating models.

     Again, this is not really surprising, because the pseudo-data are snapshots of

statistical fluctuations, and the fluctuations are not constrained by the rule used to generate

the pseudo-data. As an extreme example, it is possible (although unlikely) to generate

pseudo-data with Rule 1 but because of randomness end up with a perfect fit to Rule

2. This is a consequence of having relatively little data and therefore allowing strong fluctuations. Raising *tps* above 400 should make this kind of surprise less common.

We therefore selected the 23 sets of generating parameters underlying the minority findings. Each set was used to produce one pseudo-data matrix with 4,000 *tps*. The general VDR model ($DR_\mathrm{F}=\{1,2,3\}$) and the generating VDR model ($DR_\mathrm{F}=DR_\mathrm{G}$) were then fitted to each matrix. All 46 fits gave $r^2$ values of at least .95.

According to the *AICc*,,the general VDR model was now preferred in only five of these 23 tests with *tps* = 4,000. Furthermore, across the 23 fits of the general VDR model, the fitted rule probabilities came slightly closer to their generating counterparts. Presumably, further increases in *tps* would ultimately eliminate the five remaining cases favouring the general VDR model.

These 46 fits also permitted examination of the effects of *tps* on reproducibility (*%ic* from Equation 5*)* and on the relationship between $log(L)_\mathrm{G}$ and $log(L)_\mathrm{F}$ expressed by *%Δ*$_\mathrm{GF}$ (Equation 6). For each of these two measures, we had 52 matched values obtained under 400 and 4,000 *tps*.  For each measure and each value of *tps*, we split the general VDR fits from those where $DR_\mathrm{F}= DR_\mathrm{G}$ (one-rule and two-rule models). This produced eight subsets of data. (Wilcoxon paired data tests showed that these two classes of data could not be combined.)

First, we evaluated the effect of *tps* on *%ic*: does it change the consistency of the fits?  Wilcoxon paired data tests showed no significant effect.  For fits of the general VDR model, the Wilcoxon $V$ was 221, $p$=.258; for fits of the model with $DR_\mathrm{F}=DR_\mathrm{G}$ , $V$=120, $p$=.165.

In contrast, *tps* had a marked effect on  *%Δ*$_\mathrm{GF}$, as we expected. For fits of the general VDR model, Wilcoxon $V$=3, $p$=1.49×10$^{-7}$; for fits of the model with $DR_\mathrm{F}=DR_\mathrm{G}$ , $V$=30, $p$=6.03×10$^{-5}$. An increased *tps* lowered the mean of  *%Δ*$_\mathrm{GF}$ in both comparisons. Therefore, as *tps* increased, the fitted conditional probabilities of response came closer to the generated conditional proportions.

    *Fits of special case models to VDR-generated pseudo-data.* We can straightforwardly confirm that the VDR model is more general than SDT and CDST models.  To each of the 70 VDR-generated matrices with 400 *tps*, we fitted the SDT-EV and SDT-UV models and the general CSDT-EV and CSDT-UV models with $DR_F$={1, 2, 3}. Since the CSDT-UV model with $DR_F= DR_G$ might possibly fit better than the general CSDT-UV model to matrices generated under one-rule or two-rule VDR models, we also fitted the appropriate restricted CSDT-UV model to such matrices. This gave a total of 350 (2*70+2*70+70)  BMCMC test fits. (Recall that for SDT models, Rules 1, 2, and 3 yield identical results.)

    Table 1 gives the results. The rows are broken down by class of SDT or CSDT model. The $r^2$ values indicate that the fits were generally not as tight as those with the VDR model. However, the unequal variance forms of the SDT and CSDT models fared better than their equal variance partners. The *AICc* preferred the VDR model with $DR_G$  = $DR_F$ in the vast majority (259) of the 280 comparisons.

_______________________

Table 1 about here

_______________________

    In the other 11 comparisons, presumably the VDR model gave results close to what could be obtained via SDT or CSDT, but the size of the pseudo-experiment was not large enough to distinguish the models. To test this presumption, the generating parameters for the 11 anomalous cases were used to produce new pseudo-data matrices with 4,000 *tps*. We

fitted the models in question to the new pseudo-data. All 11 previously unexpected comparisons now reversed in favour of the general VDR model.

### SDT- and CSDT-generated pseudo-data

The general VDR model gave fits with high $r^2$ values to pseudo-data generated by DR models with all possible classes of decision processes. This observation leads to a final sanity check: Is the general VDR model so powerful that it can give superior fits to SDT- and CSDT-generated pseudo-data? To answer this question, we generated pseudo-data with the equal-variance and the unequal-variance versions of the SDT and CSDT models. For the CSDT models, we allowed only the general rule set {1,2,3}.

Ten pseudo-data matrices were formed with 400 *tps* under each of the four generating conditions. The generating SDT or CSDT model and the general VDR model were fitted to each matrix; the *AICc* selected the better fitting model in each pair. Table 2 summarizes the results.

*Fits to SDT-generated matrices.* The first two rows of Table 2 show the outcomes of fits of the general VDR and generating SDT models to SDT-generated data. Both classes of model generally gave fits as tight as those reported above to VDR-generated pseudo-data. The few mild departures occurred in three SDT and three VDR fits to SDT-EV-generated matrices. Values of $r^2$ fell to .94. All 14 other fits gave $r^2$ values of at least .95. In the six mildly weaker fits, one value of *rmsd* reached .02. Four values of *K-L* exceeded 0.05, the largest being 0.09. Most importantly, in each of the 20 pairs of fits, the *AICc* chose the SDT model over the general VDR model.

_______________________

Table 2 about here

_______________________

Since fitting a SDT model to data does not require numerical integration, we can test the hypothesis that spurious minima from control of Romberg step-size often trap the BMCMC optimisation.  Under this hypothesis, we expect BMCMC to converge much more closely to the global optimum for SDT fits vs. VDR fits, because numerical integration is not needed for SDT evaluations. Therefore, across multiple fits, *%ic* should be smaller for the SDDT fits.  This is generally what we see: 10 of 10 fits of SDT-UV models had a smaller *%ic* than the corresponding fit with a VDR model (binomial *p* < 0.0001), and likewise for 9 of 10 fits with a SDT-EV (p binomial *p* < 0.001).  We consider this a successful prediction.

*Fits to CSDT-generated matrices*. Fits to the ten CSDT_EV-generated matrices were less successful. (See the third row of Table 2.) In five cases, both the CSDT and the VDR model gave strong fits, and the *AICcc* always deemed the CSDT-EV model as the better. For the other five pseudo-data matrices, the CSDT-EV model did not fit as well. The $r^2$ values dropped as low as .89, *rmsd* always was .02 or more, and *K-L* climbed as high as 0.38. In contrast, the VDR fits were very good. Consequently, the *AICc* selected the general VDR model as better than the CSDT-EV model in these five comparisons.

Once again, it seemed likely that the relatively low value of *tps* underlay these five irregular outcomes. Using the generating parameters for the five matrices in question, we simulated new experiments with 4,000 *tps*. We then fitted the CSDT-EV and general VDR models to the enlarged matrices. All fits were excellent, with $r^2$ values of .96 or better. The *AICc* now chose the CSDT-EV model over the general VDR model.

Finally, the VDR and CSDT-UV models gave the usual tight fits to pseudo-data generated by the CSDT-UV model. (See last row of Table 2.)  In nine instances, the *AICc* chose the  CSDT-UV model as superior to the general VDR model. As before, raising *tps* to 4,000 reversed the previous outcome and the AICc now preferred CSDT-UV.

In the VDR fits to SDT- and CSDT-generated pseudo-data, *all* criterion and representation variances were nonzero, although some occasionally came close to zero. Despite this, the general VDR model gave surprisingly good fits to the SDT- and CSDT-generated pseudo-data. Across all comparisons between the generating model and the VDR model, the per cent difference between the *AICc* values was small (mean=0.56, s.d.=0.03) and positively skewed.

## Discussion

Our findings raise two main questions. First, the how can the sample size (*tps*) be determined for an experiment aimed at a choice between models? Second, what is an efficient way to select between the many possible models for rating data?

*Choosing a sample size for model selection*. In our simulations, the *AICc* sometimes yielded an unexpected choice between models fitted to the same pseudo-data. We found that we could reverse such choices by increasing trials per stimulus (*tps*) and thus the sample size. Clearly, an *AICc* selection can change from sample to sample with increased sample size.

There should be a finite critical sample size (*css*) beyond which more observations would not change model selection. Obviously, *css* would depend on the models undergoing comparison; in particular, it should increase with the number of model parameters. We cannot provide an algorithm for computing the *css*. We can, however, offer a heuristic: *simulation before experimentation*.

Suppose that an experimenter sets out to choose between two models A and B. She should simulate each one at different sample sizes and fit each model to each set of pseudo-data. She should look for a sample size beyond which A remains the better choice for explaining A-generated pseudo-data and B remains the better choice for explaining B-generated pseudo-data. This gives an estimate of the *css*. It also sets a lower bound to the sample size of the planned experiment. This heuristic also can be used post hoc to test the stability of previously reported model comparisons.

To choose reliably between different models, data sets an order of magnitude larger than those commonly used may well be required.  To acquire that much data, we may need to move away from experiments that focus on the individual observer and develop new experimental designs suitable for large numbers of subjects.

*Selecting the best model for rating data*.  There are seven different classes of VDR models, depending on the choice of decision rule or rules, and it includes SDT and CSDT rating models as special cases. Given a matrix of rating data, is there an efficient procedure for selecting the best model from this formidable array of possibilities?

Every model in this array assumes that successive responses are independent. Therefore, an initial step must verify that the data meet this assumption. This requires calculating a partial autocorrelation function on the data. One or more nonzero values are bound to occur; small values can be tolerated if they fall around or below the 95 per cent confidence limit and at lags 1 and 2. Model fitting can then proceed. One or more autocorrelations beyond this limit make model fitting a dubious exercise.

If response dependencies are small, our results suggest a procedure for model fitting. First, fit the general VDR model ($DR$={1, 2, 3}) and the SDT-UV model to the observations. These two models gave good fits to a variety of pseudo-data. If the *AICc* prefers the SDT-UV model, criterion variances must be zero. Then the SDT-EV model should then be tried and tested against the SDT-UV model.

If the *AICc* prefers the general VDR model, however, a further step may be needed. Our results show that the fitted rule probabilities will give some clues to the observer's decision procedure. To pursue these clues, fit the appropriate dual- or single-rule VDR model to the data and compare the two VDR models through the *AICc*.

Note that the VDR and SDT models embody the standard assumption that the representation variance is nonzero.  To test and confirm this, fit the general CSDT-EV and CSDT-UV models to the data and compare it to the model previously selected as best.  If all representation variances are nonzero, CSDT should lose in both comparisons.

*An example.* We illustrate this procedure with rating data from two experiments by Schouten and van Hessen (1998). They elicited so-called 'magnitude judgments' of a) the loudness of ten 1-kHz tones differing in intensity and b) the position of 11 synthetic speech stimuli on a one-dimensional continuum of Dutch /pak/-/tak/-/kak/ syllables. On each trial, the subject responded to a single randomly chosen stimulus by positioning a mouse-driven pointer along a horizontal bar presented on a display. For the intensity experiment, the ends of the bar were labelled 'soft' and loud'. For the speech experiment, the ends were labelled 'p' and 'k'.

Before each block of 100 trials, two extreme reference stimuli were presented. All experimental stimuli were strictly inside the range of the reference stimuli. Markers slightly inside the ends of the response bar represented the reference stimuli. The same four subjects participated in both experiments. With 400 trials per stimulus per subject per experiment, eight unusually large data sets resulted.

Schouten and van Hessen (1998) visually inspected each subject's histogram of responses to each stimulus. The authors claimed that every histogram tended to be Gaussian. Furthermore, each of the four sets of histograms from the intensity experiment seemed to show equal variances across stimuli. Schouten and van Hessen took this as support for the SDT-EV model.  In contrast, the histograms from the speech experiment displayed unequal variances.

Pastore and Macmillan (2003) pointed out that Schouten and van Hessen (1998) had assumed that each subject's responses expressed an interval scale. Pastore and Macmillan reanalysed the data with the SDT rating scale model that assumes only ordinal properties for the subject's responses (see Macmillan and Creelman, 2005, ch. 3). After binning the data, Pastore and Macmillan plotted one-step and two-step receiver operating characteristic (ROC) curves for each subject and each possible stimulus condition.  Chi-square tests revealed that over 75% of the ROCs from the intensity experiment and about 65% of those from the speech study conformed to the SDT model. Visual inspection of the rest suggested that most followed Luce's (1963) low threshold model.

This reanalysis undercut Schouten and van Hessen's (1998) assertion that all their response histograms were Gaussian. Fitting ROC curves, however, is a relatively weak test. Each plot uses only some fraction of a data set. In a stronger test, we followed our recommended procedure for treating rating data.

First, partial autocorrelations were obtained through lag 10 across each of Schouten and van Hessen's (1998) eight raw data sets.  A data set underwent model fitting if it showed small enough response dependencies. We placed the responses in 10 to 13 bins. This avoided a large number of cells with small or zero values in the resulting S-R matrices, while permitting sufficient variation across cells.  Then we fitted the VDR and SDT-UV models to the data set.  Three different starting points were used for each fit. The one yielding the highest value of $log(L)_F$ was accepted.

*Intensity data.* One intensity data set yielded a lag-1 autocorrelation of  .037 just above the upper 95 per cent confidence level of .031. These data underwent model fitting. Stronger partial autocorrelations, however, marked the rest of Schouten and van Hessen's (1998) intensity data. At lag 1, two intensity data sets each had a large partial autocorrelation (.203 and .093, respectively). The remaining data set showed substantial coefficients out through lag 5 (.117, .083, .064, .045, and .052, in lag sequence) that breached the upper 95 per cent confidence level. Since these three intensity data sets clearly violate a basic assumption of detection theory models, they were dismissed from further analysis. Furthermore, they cannot be offered as support for SDT, contrary to Schouten and van Hessen's claims.

Both the general VDR and the SDT-UV models fitted the one available intensity data set tightly, with $r^2$ values above .95. The *AICc* preferred the VDR model. The probabilities were highest for Rules 2 and 3 in the general VDR fit. We therefore also fitted VDR models with *DR*={2,3} and *DR*={3} to the intensity data set. The *AiCc* selected the fit with *DR*={2,3} as the best of the three,

Schouten and van Hessen (1988) contended that their intensity data conformed to the SDT-EV model. We fitted that model to the one usable intensity data set. The $r^2$ value reached .94, signalling a good fit. The *AICc*, however, chose the general VDR model and the SDT-UV model over the SDT-EV model. This contradicts Schouten and van Hessen's contention.

*Speech data.* Two speech data sets showed no significant response dependencies whatsoever. The other two each produced one autocorrelation at lag 1 or 2 that just exceeded the upper 95 per cent confidence level. Such small autocorrelations are to be expected occasionally and at worst only weakly violate the assumption of response independence. Accordingly, all four sets of speech data underwent the planned model fitting.

The smaller response dependencies in the speech as against the intensity data may well reflect extensive experience in judging speech sounds. Criterion-setting theory (Treisman and Williams, 1987) provides for such an effect. Experimenters who want to minimize response dependencies should consider synthetic speech stimuli.

^Both the general VDR and SDT-UV models always gave fits with $r^2$ values above . 95. The *AICc* preferred the SDT-UV model to the VDR model for explaining one of the four speech data sets.  This data set supports Schouten and van Hessen's (1998) claim that the speech data obey the SDT-unequal-variance model.  Our simulation results reported above suggest that these experiments (with *tps*=400) are not large enough to allow completely reliable comparisons between models, so one shouldn't read too much into a single case.

For the other three speech data sets, the general VDR model with its nonzero criterion variances provided superior fits. The VDR decision rule probabilities made Rule 2 a heavy favourite for explaining two of these data sets. We fitted appropriate single- and dual-rule models to these data. However, the *AICc* still chose the general VDR model as best for both data sets.

The rule probabilities for the remaining data set favoured Rules 1 and 2. Appropriate dual-rule and single-rule models were fitted to this matrix. The *AICc*  chose the single-rule decision procedure DR={1} over all other candidates. The same subject had produced this speech data set and the single usable intensity data set. The preferred decision procedure diffe22red between experiments.

*Test for nonzero representation variances.* For this test, we fitted the CSDT-EV and CSDR-UV models with $DR_F$={1, 2, 3} to all five data sets. The CSDT-EV model gave the poorest fits. Three of the five values of $r^2$ fell below .90. The other measures of goodness of fit told the same story. In contrast, four of the five fits of the CSDT-UV model yielded $r^2$ values of at least .95. Nonetheless, the *AICc* always chose the general VDR model over the CSDT models.

In addition, the *AICc* chose the SDT-EV model over the CSDT-EV model in three of the five comparisons. The SDT-UV model prevailed in four of the five comparisons against the CSDT-UV model. The relatively weak showing by the CSDT models supports the fundamental assumption in psychophysics of trial-to-trial variation in the representations of a given stimulus.

## Conclusion

We have built a variable decision rule (VDR) rating model that mathematically embodies all of Klauer and Kellen's rating scale decision procedures. Despite its mathematical complexity and non-linearity, the VDR model can be fitted to data fairly straightforwardly with the Bootstrap Markov Chain Monte Carlo (BMCMC) algorithm.

The fits are good, matching the data accurately, even though our implementation of the

mathematics generates shallow, spurious minima. (The minima seem to arise spuriously

from sudden jumps in likelihood due to discrete changes in the Romberg integration step.)

Our simulations show that a large experiment is necessary to determine which decision rule(s) were used to generate the data.  At 400 trials per stimulus (*tps*), some information can be obtained, but the statistical fluctuations are still large enough to prevent distinct and reliable estimates of rule probabilities. Better results are obtained at 4000 *tps*, but actual studies this large may require new experimental approaches.

Using the Akaike Information Criterion (corrected) (*AICc*), we can choose between

VDR, SDT, and CSDT models fitted to the same data. Reliable choices, however, may also require large experiments, since the *AICc* is sensitive to sample size. Large experiments are costly, so we propose a simulation technique for deciding how many observations an experiment requires in order to reach particular goals.

Finally, we reanalysed Schouten and van Hessen's (1988) rating data for tonal intensity and for a synthetic speech continuum. Inter-response dependencies made most of the intensity data unusable. The remaining usable data sets were mostly better fitted by the general VDR model than by the claimed classical signal detection theory model. Fitted rule probabilities suggest that different subjects used different decision procedures.

Complementary signal detection theory models did not provide good fits, supporting the fundamental assumption of representation variation.

# Author Note

Burton S. Rosner, Phonetics Laboratory, University of Oxford, 41 Wellington Square,

Oxford OX1 2JF, U. K. email: burton.rosner@phon.ox.ac.uk.

Greg Kochanski, Google, Inc. email: gpk320@gmail.com

We thank John Hainsworth for helpful comments. M. E. H. Schouten and A. J. van Hessen kindly provided the original the data from their experiments. We also are indebted to R. E. Pastore and N. A. Macmillan for complete files of their reanalysis of those data

Table 1.

Fits of different special case rating scale models to simulations (400 trials per stimulus) of the general variable decision rule model (VDR$_{123}$).

| n$_G$ | VDR$_{123:}$ | Y | Y: | $n$(VDR$_{123}$ f Y) |
|---|---|---|---|---|
| | $n$ ($r2 \geq .95$) | | $n$($r^2 \geq .95$) | |
| 70 | 69 | SDT-EV | 0 | 70 |
| 70 | 69 | SDT-UV | 14 | 66[a] |
| 70 | 69 | CSDT-EV$_{123}$ | 3 | 69[a] |
| 70 | 69 | CSDT-UV$_{123}$ | 29 | 64[a] |

Notes:  n$_G$ =number of VDR$_{123}$ simulations;  VD$_{R123}$: $n$($r2 \geq .95$)=number of fits of VDR$_{123}$ with $r^2$ at least .95;  Y =class of fitted model; Y: $n_Y$($r^2 \geq .95$)=number of fits of model Y with $r^2$ at least .95; $n$(VDR$_{123}$ f Y) =number of preferences for VDR$_{123}$ over model Y; SDT-EV =equal-variance SDT model;  SDT-UV =unequal-variance SDT model; CSDT-EV =equal-variance complementary SDT model with all decision rules; CSDT-UV =unequal-variance complementary SDT model with all decision rules.

[a]all  Y f  VDR$_{123}$ reversed with 4,000 trials/stimulus.

Table 2.

 Fits of generating and $VDR_{123}$ rating scale models to simulations (400 trials per stimulus) of SDT and CSDT rating scale models.

________________________________________________________________

| X | X: $n(r2{\geq}.95)$ | $VDR_{123}$: $n(r^2 \geq .95)$ | $n(X\ f\ VDR_{123})$ |
|---|---|---|---|
| SDT-EV | 9 | 8 | 10 |
| SDT-UV | 10 | 10 | 10 |
| CSDT-EV | 5 | 10 | 5[a] |
| CSDT-UV | 10 | 10 | 9[a] |

________________________________________________________________

Notes:

X =generating model and fitted model; $n_X(r2{\geq}.95)$=number of fits of model X with $r^2$ at least .95; $n_{VDR_{123}}(r^2 \geq .95)$ =number of fits of general VDR model with $r^2$ at least .95;

$n(X\ f\ VDR_{123})$ =number of preferences for model X over general VDR model; SDT-EV =equal-variance SDT model;  SDT-UV =unequal-variance SDT model; CSDT-EV =equal-variance complementary SDT model with all decision rules; CSDT-UV =unequal-variance complementary SDT model with all decision rules.

[a]all  VDR f  *X* reversed with 4,000 trials/stimulus.

**Footnotes**

[1] ransim_a.cpp. All code described here is part of this PDF, as an attachment.  The paper-specific code is in Rosner_Kochanski_2016.tgz (in gzipped tar format), and the libraries we used (including the BMCMC code) is in 2015-12-02_speechresearch.tgz (again, gzipped tar format; note that we removed some example/test files to keep the size under arXiv.org's 6 MB limit).

[2] The optimisation program has three modules; two are written in Python (bsr_analysis.py and bsr_analysis_guts.py) and the other (lawcjcrk.cpp) is a C++ module.

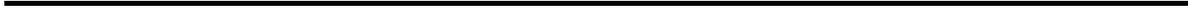